\documentclass[12pt,preprint]{aastex}

\def\co{C$^{18}$O \ }

\def\nht{NH$_3$ \ }

\def\msun{M$_\odot$}

\slugcomment{To appear in the Astrophysical Journal}

\shorttitle{The Nature of Dense Cores}
\shortauthors{C.J. Lada,  A.A. Muench, J. Rathborne, J.F. Alves \& M. Lombardi}

\begin{document}



\title{The Nature of the Dense Core Population in the Pipe Nebula:\\
Thermal Cores Under Pressure}

\author{Charles ~J. Lada\altaffilmark{1},
  A.A. Muench\altaffilmark{1}, J. 
  Rathborne\altaffilmark{1}, Jo\~ao~F. Alves\altaffilmark{2} \& M. Lombardi\altaffilmark{3,4}}

\altaffiltext{1}{Harvard-Smithsonian Center for Astrophysics, 60 Garden Street, 
Cambridge, MA 02138, USA; clada@cfa.harvard.edu}
\altaffiltext{2}{Calar Alto Observatory, Centro Astron\'{o}mico Hispano
Alem\'{a}n, c/Jes\'{u}s Durb\'{a}n Rem\'{o}n 2-2, 04004
Almeria, Spain. \mbox{jalves@caha.es}}
\altaffiltext{3}{European Southern Observatory, Karl-Schwarzschild-Str. 2, 85748 
Garching, Germany}
\altaffiltext{4}{University of Milan, Department of Physics, via Celoria 16, 20133
Milan Italy}

\begin{abstract}

In this paper we present the results of a systematic investigation of an entire population
of dust cores within a single molecular cloud.  This population of predominately starless
cores was previously identified in an infrared extinction survey of the Pipe Nebula, a
nearby molecular cloud with a well-established distance but negligible star forming
activity.  Analysis of extinction data shows the cores to be dense objects characterized by
a narrow range of density with a median value of $n(H_2) = 7 \times 10^3$ cm$^{-3}$,
assuming a normal gas-to-dust ratio.  Analysis of C$^{18}$O and NH$_3$ molecular-line
observations reveals very narrow lines.  The non-thermal velocity dispersions measured in
both these tracers are found to be subsonic for the large majority of the cores and show no
correlation with core mass (or size).  The bulk gas motions are thus acoustic in nature and
thermally dominated.  Thermal pressure is thus the dominate source of internal gas pressure
and support for most of the core population.  The total internal gas pressures of the cores
are found to be roughly independent of core mass over the entire (0.2-20 \msun) range of the
core mass function (CMF) indicating that the cores are in pressure equilibrium with an
external source of pressure.  This external pressure is most likely provided by the weight
of the surrounding Pipe cloud within which the cores are embedded.  Most of the cores appear
to be pressure confined, gravitationally unbound entities whose nature, structure and future
evolution are determined by only a few physical factors which include self-gravity, the
fundamental processes of thermal physics (i.e., heating and cooling) and the simple
requirement of pressure equilibrium with the surrounding environment.  The observed core
properties likely constitute the initial conditions for star formation in dense gas.  The
entire core population is found to be characterized by a single critical Bonnor-Ebert mass
of $\approx$ 2 \msun.  This mass coincides with the characteristic mass of the Pipe CMF
indicating that most cores formed in the cloud are near critical stability.  This suggests
that the mass function of cores (and ultimately the stellar IMF) has its origin in the
physical process of thermal fragmentation in a pressurized medium.

\end{abstract}


\section{Introduction} \label{sec:introduction}

The stellar initial mass function (IMF) is one of the most fundamental distributions in
astrophysics.  Its origin is one of the most critical but least understood aspects of the
star forming process and is perhaps the most fundamental unsolved problem of star
formation (e.g., Corbelli, Palla \& Zinnecker 2005; Bonnell, Larson \& Zinnecker
2007).  Stars form in the dense cores of molecular clouds, but little is understood
about the detailed physical properties of these cores prior to star formation and even
less is understood about their origin.  However, both these issues are critically linked
to the origin of the IMF.

Lack of even a rudimentary understanding of these issues stems primarily from a general
lack of empirical knowledge of earliest stages of dense core evolution.  This state of
affairs is a result of an absence of molecular clouds identified in a sufficiently early
evolutionary state that little or no star forming activity is present within them.  This
is often taken as evidence that the onset of star formation in molecular clouds is
extremely rapid and that clouds in the earliest evolutionary states must be exceedingly
rare.  Clearly the identification and detailed study of such a cloud and its dense core
population would be of great interest. 

We have identified the visually prominent dark cloud known as the Pipe Nebula as one of
the best candidates for an extremely young molecular cloud complex.  At a distance of 130
pc (Lombardi, Alves \& Lada 2006) this cloud is one of the nearest objects of its mass
($\sim 10^4$ \msun) and size ($\sim$ 3 $\times$ 14 pc) to the sun.  The cloud exhibits
very little evidence of star formation activity and this is probably reflected in the fact
that prior to 2006 only one paper in the literature was devoted to its study.  That paper,
a CO survey by Onishi et al.  (1999) showed that despite the paucity of star formation
activity the cloud was similar enough in its $^{12}$CO and $^{13}$CO emission to the
Taurus clouds that its potential to form stars was high.

\newpage
These considerations motivated us to undertake a detailed and systematic multi-wavelength
investigation of this cloud in order to quantify its overall properties.  The primary goal
of this program was to identify the complete population of its dense cores and then
determine their detailed physical nature.  The study of the core population in a single
cloud insures that all the cores are at the same distance, removing one of the largest
sources of systematic uncertainty in relative comparisons of core physical properties.  The
proximity of the Pipe cloud is critical to enable even the smallest and lowest mass cores to
be detected and resolved.  In the first step of this investigation we obtained a wide-field,
near-infrared extinction map of the entire 6$^o$ x 8$^o$ extent of the Pipe region and
quantified the basic structure and properties of the cloud (Lombardi, Alves \& Lada 2006).
Using that data we then identified its core population and determined their masses (Alves,
Lombardi, \& Lada 2007).  The dust extinction measurements enabled robust determination of
core masses over a large dynamic range (0.2--20 \msun) with more statistically significant
sampling to lower masses than previously achieved using different techniques in other
clouds.  The Pipe cores appear to be a pristine population of starless objects whose mass
spectrum displays an overall shape (and width) nearly identical to that of the stellar IMF,
but offset in mass by a factor of 3.  These observations suggested that the core mass
function (CMF) of the Pipe Nebula is the direct precursor to the stellar IMF, once it is
modified by a constant star formation efficiency (SFE) of $\sim$ 30\%.

Determination of the detailed physical properties of such a population of (mostly
starless) dense cores can not only provide important information about the initial
conditions of star formation within the individual cores, but also potentially new
insights concerning the nature and the origin of the CMF and perhaps even the IMF itself.
Our measurements of dust extinction can provide robust determinations of fundamental core
properties such as the sizes, masses and densities.  To obtain a complete description of
basic core properties however requires direct observations of the gas within the cores.
Observations of molecular lines can provide critical information about both the overall
kinematics and the internal dynamical states of dense cores.  This information in turn
enables determinations of the boundedness and stabilities of the cores as well as the
internal pressures characterizing them.  Therefore in an effort to obtain a more complete
physical description of the core population in the Pipe cloud we augmented our extinction
survey with two molecular emission line surveys of the core population.  The first survey
was a C$^{18}$O survey of a large fraction (2/3) of the cores designed to provide
information about the kinematics and dynamical nature of as complete as sample of the
cores as possible (Muench et al.  2007).  This was followed by a directed NH$_3$ study of
a smaller but significant sample of the core population to measure the conditions in the
densest material (Rathborne et al.  2007).  In this paper we combine the results of the
molecular-line surveys with the extinction observations to perform a detailed examination
of the basic physical properties of this important core population.

\section{Results and Analysis}

\subsection{Dense Cores} \label{sec:extinctionobs}

Alves, Lombardi and Lada (2007; hereafter ALL07) used a wavelet decomposition technique to
identify 159 cores in the Pipe cloud from the wide-field extinction map of Lombardi, Alves
\& Lada (2006; hereafter LAL06).  The wavelet decomposition effectively acts as a low pass
filter to remove a relatively extended smoothly varying background from the extinction
image.  This results in the construction of a "cores only" image of the small-scale ($l
\leq $ 0.3 pc) spatial structure in the cloud.  Visual inspection of this image showed a
population of well defined and well separated extinction peaks or cores.  An automated
two-dimensional algorithm (Clumpfind2d; Williams, De Geus \& Blitz 1994) was used to
systematically identify and extract cores in an objective manner.  The detection threshold
was set at $A_V = 1.2$ magnitudes, which is three times the rms variation in extinction
over extended regions of the image nearby but off the cloud and six times the measured rms
uncertainty in individual pixels of the original extinction image (LAL06).  The input
parameters for the algorithm were adjusted to minimize and eliminate spurious core
identifications.  This requirement was met by demanding visual verification of each core
extracted.  The primary advantage of using this technique is to produce a systematic and
objective measurement of the area, $A$, of each core which can then be converted to a
radius via $R = \sqrt{A/\pi}$.  This radius corresponds to the outer edge of the core.
The mass of a core is then calculated by integrating the (background subtracted) dust
column density over the area of the core and multiplying by an assumed gas-to-dust ratio
(in this case a ratio of 100).

Because the uncertainties in the extinction measurements are extremely small (on
the order of a few \%), the primary source of systematic uncertainty in the
derived masses (apart from the uncertainty in the distance to the cloud) is the
uncertainty in the derived area of an extracted core.  To estimate this
uncertainty we performed a number of core extractions varying the input
parameters, including the threshold extinction while relaxing somewhat the visual
verification constraint.  We then compared the results for individual cores
identified in common.  From these experiments we estimate the typical uncertainty
in core area (and mass) to be between 10-30\%.  Thus the extracted core radii
and derived masses appear to be quite robust.

With good masses and radii in hand we can then calculate the
mean densities of each core, assuming spherical symmetry, from $\rho = 3M/(4 \pi
R^3)$.  The corresponding volume number density is $n(H_2) = \rho/\mu m_H$ where
$\mu$ is the molecular weight (2.34, for a molecular hydrogen gas) and $m_H$ is
the mass of a proton.  In table A1 we list the radii, masses and densities
determined for the cores in the Pipe nebula.

Figure \ref{coredensity} displays the frequency distribution of core number densities in
the Pipe cloud.  The core density distribution is characterized by a well defined and
relatively narrow peak.  The mean core density is found to be 7.3 $\times$ 10$^3$
cm$^{-3}$.  The dispersion in density is $2 \times 10^3$ cm$^{-3}$.  The distribution is
not perfectly symmetric about this mean and displays a significant tail to higher
densities.  The median density of the cores is found to be 7.1 $\times$ 10$^3$ cm$^{-3}$.
As discussed by Rathborne et al.  (2007), this is consistent with the 
NH$_3$ emission exhibited by the core population in the Pipe cloud.  Although the typical
density of the Pipe cores is somewhat lower than that of a few $\times$ 10$^4$ cm$^{-3}$
usually attributed to cores in other clouds, (e.g., Jijina et al.  1999), the core
population identified by ALL07 and investigated here is a population of dense cores.

\subsection{Thermally Dominated Cores}
\label{sec:observations}

Molecular emission-lines provide important information about the kinematics and internal
dynamics of molecular clouds.  A little over a quarter century ago Larson (1981) used
observations of CO to establish that molecular clouds appear to satisfy three general
relations, now known as Larson's Laws.  The first of these laws expresses a relation
between the size of a cloud and its velocity dispersion, namely $\sigma \sim R^{0.5}$.
On the scales of dense cores $\sigma$ is determined directly from the linewidth.  The
second law was that of approximate virial equilibrium, $ (\sigma^2R)/(GM) \sim 1$.  The
third law, which follows from the first two, relates the sizes and masses of the clouds
and indicates that clouds are characterized by constant column density, that is, $M \sim
R^{2}$.  These laws are widely interpreted to indicate that the physical state of
molecular clouds is best described by a common hierarchy of (supersonic) turbulent motions
under the influence of gravity (Larson 1981).

Figure \ref{ntsigmas} plots the one dimensional non-thermal velocity dispersions,
$\sigma_{NT}$, in the individual Pipe Nebula cores versus core mass for both \co and \nht
observations.  These velocity dispersions were calculated from the molecular linewidths
assuming a gas kinetic temperature of 10 K.{\footnote{ Rathborne et al. (2007)  were able to
estimate kinetic temperatures for 12 cores and found that higher mass cores, with the
strongest \nht detections were characterized by temperatures $\sim$ 10 ($\pm$ 1) K while
lower mass cores with weaker detections were characterized by temperatures of $\sim$ 13
($\pm$ 3) K.  For the purposes of this paper we conservatively adopt a single kinetic
temperature of 10 K for all the cores.}  There is no correlation between non-thermal
velocity dispersion and core mass.  Moreover, the vast majority of these velocity
dispersions have magnitudes less than the speed of sound, 0.2 km/s, in a 10 K molecular
gas.  The non-thermal motions in these cores are thus subsonic across the entire spectrum
of core masses in the cloud, a spectrum which covers nearly two orders of magnitude in
mass.  This suggests that the gas motions in the cores are acoustic and thermally
dominated.  A very important consequence of this finding is that the cores in the Pipe
Nebula must evolve on acoustic, thus relatively slow timescales.  The typical core sound
crossing time is about 10$^6$ yrs.

The results of figure \ref{ntsigmas} coupled with the absence of a size-linewidth relation
in both the \co and \nht lines for the Pipe core population (Muench et al 2007; Rathborne
et al 2007) indicates that this population violates Larson's Laws for turbulent molecular
clouds.  If the non-thermal motions in the Pipe cores are due to turbulence, this
turbulence is of a different nature than that which is typically described by Larson's
Laws.  It is subsonic, rather than supersonic and is characterized by a scale which is
smaller than the core size and thus it is more likely micro-turbulent rather than
macro-turbulent in nature.

To further investigate the possibility of thermally dominated motions in the Pipe cores,
we consider the ratio of thermal to non-thermal pressure i.e., $R_p = a^2/\sigma_{NT}^2$,
for each core; here $a$ is the one-dimensional isothermal sound speed in a 10 K gas.  In
figure \ref{Rpressures} we plot $R_p$ vs mass.  For more than 2/3 of the cores measured in
CO and 80\% of the cores measured in \nht $R_p$ $>$ 1 and the thermal pressure clearly
exceeds the non-thermal pressure.  Indeed, in the most extreme cases, the thermal pressure
is 1-2 orders of magnitude higher than the non-thermal pressure.  For three of these cores
the observed \nht linewidths were indistinguishable from purely thermally broadened line
profiles in a 10 K gas and lower limits to $R_p$ are plotted.\footnote{These three cores
could also be characterized by $T_k < 10$ K, in which case $R_p$ could be somewhat lower
than plotted.}  For 90\% of the CO measurements and all the \nht measurements $R_P > 0.5$.
Therefore thermal pressure is a significant if not the dominant source of internal gas
pressure for essentially the entire core population.

We note here that our calculated values of $R_p$ are likely underestimates to the true
values for many sources.  This is due to two reasons.  First, our \nht observations
show that the cores in the Pipe can have temperatures as high as 12 - 15 K (Rathborne et
al.  2007) and thus our assumption of 10 K for the typical gas temperature may
underestimate the actual gas temperatures for some fraction of the objects.  Second, for
most cores the non-thermal pressures were calculated from the \co observations and this
could result in an overestimate of the non-thermal pressure since the \co observations
sample both core and more extended inter-core or background gas.  This is particularly
true for the lower mass cores where as much as half or more of the total line-of-sight
column density sampled by the \co observations arises outside the core.  Indeed, for 25
cores (excluding the 3 sources with lower limits) where both \nht and \co are observed,
the non-thermal pressure derived from the CO data is on average a factor of 4 larger than
that measured by the ammonia line.

The results described here are consistent with and confirm those derived by Barranco \&
Goodman (1998) and Goodman et al.  (1998) who studied four dense cores in four different
clouds and, similar to the findings presented here, found the cores to be characterized by
subsonic non-thermal motions.  They argued that dense cores represented regions of
``coherence'' in otherwise turbulent molecular clouds.  Apparently the physical conditions
which characterize the population of cores in the Pipe cloud may be typical of dense cores
in general, at least in regions of low mass star formation.

\vskip 0.2in

\subsection{Pressure Confined Cores} \label{sec:results}
 
\subsubsection{Gravitational Binding} 
 
In this section we investigate the integrity of the cores as 
persistent, coherent entities. First, we consider the gravitational
binding of the cores. For each core we calculate the three dimensional
velocity dispersion including both thermal and non-thermal contributions,
i.e.,

$$\sigma_{3D} = \sqrt{3 a^2 + 3\sigma_{NT}^2}$$

\noindent
and we compare this to the escape velocity of a spherical core with
the same mass ($M$) and size ($R$), i.e, 

$$V_{esc} = \sqrt{2GM/R}$$

\noindent 
In figure \ref{gbind} we plot the ratio of $\sigma_{3D}$ to $V_{esc}$ against the log
of the core mass.  This ratio of gas velocity dispersion to escape speed is well
correlated with core mass.  The majority of the cores appear to be unbound
gravitationally.  Thus these cores appear to violate the second of Larson's laws,
that of gravitationally bound objects.  The threshold between bound and unbound cores
occurs at about 2-3 \msun.  It is interesting that this mass similar to the
mass at which the core mass function (CMF) breaks from a single power-law form before
reaching its peak (ALL07).  If they form stars, the bound cores will likely produce
stars that populate the Salpeter (1955), power-law portion of the stellar IMF.

\subsubsection{Internal Pressures and Pressure Confinement}

What is the nature of the unbound cores? To investigate this further we
calculated the average total (thermal $+$ non-thermal) internal gas pressures 
of each of the Pipe cores, i.e., 

\begin{equation}
P({\rm total}) = P({\rm T}) + P({\rm NT}) = \rho(a^2 + \sigma_{NT}^2)
\end{equation}

\noindent 
where $\sigma_{NT}$ is the one-dimensional, non-thermal velocity dispersion and
$\rho$ is the mean density of a core calculated from the mass and size derived for it
from the extinction data.  In figure \ref{totalpressure} we plot the total pressure,
$P({\rm total})/k$, vs core mass.  The plot displays significant scatter (about a
factor of 3 in pressure independent of core mass) and shows that the internal core
pressure is not a particularly strong function of mass.  Both the core pressure and
spread in this pressure are surprisingly similar over the entire range of core mass.
The facts that the cores are spread out over the entire 14 pc length of the cloud and
yet have very similar internal pressures, independent of whether they are 0.2-0.3
solar mass cores or 10-20 solar mass cores, cannot be a coincidence.  These facts
strongly suggest that the cores are in pressure equilibrium with an external source
of pressure which encompasses all the cores and which physically communicates and
sets their surface pressures.

Figure \ref{mvsr} shows the mass-radius relation for the Pipe cores.  There is a
relatively tight correlation between mass (M) and radius (R).  A linear least-squares fit
to the data gives $M \sim R^{2.56 \pm 0.05}$.  As mentioned earlier, for clouds that obey
Larson's Laws we would expect constant column density and $M \sim R^{2}$.  However, for a
core population characterized by both a constant internal thermal pressure and a constant
kinetic temperature, we would expect the core volume density to also be constant and this
would result in a mass-radius relation of the form $M \sim R^{3}$.  The observed relation
is closer to the expectations of constant volume density than constant column density.
The fact that the data points all lie clearly above the sensitivity threshold 
for the observations also demonstrates that the narrow range in density derived 
for the cores in Figure 1 and manifest here is not an artifact of observational selection.
This finding provides further evidence in support of the notion that the internal
pressures of the Pipe cores are thermally dominated and are all characterized by
essentially the same surface pressure.

An often used metric to evaluate the relative importance of the gravitational and kinetic
energies of a dense core is the virial parameter:  $\alpha = (5\sigma^2R)/(GM)$ (McKee
1998).  In their theoretical study of pressure confined cores in magnetized clouds,
Bertoldi \& McKee (1992) argued that for pressure- confined cores $\alpha$ should depend
on mass as $\alpha \sim M^{-2/3}$.  To illustrate this consider that for an isothermal,
constant density, pressure-confined core, $M \sim R^{3}$ as mentioned above, thus using
the definition of $\alpha$ , $\alpha \sim M^{1/3}/M \sim M^{-2/3}$.  In figure
\ref{virialparameter} we plot $\alpha$ vs mass for the Pipe core population.  There is a
strong correlation between the two parameters and for all cores $\alpha > 1$.  A
least-squares fit to the data gives $\alpha \sim M^{-0.66 \pm 0.04}$, closely matching the
expectations for an ensemble of pressure-confined cores.  The fact that $\alpha$ is not a
constant indicates again that the dense core population is in violation of Larson's Laws.
The fact that none of the cores appear  virialized suggests that the
core population is extremely young.

These considerations imply that the cores in the Pipe are pressure confined entities.
Even though most are gravitationally unbound, they are still all coherent objects
which will persist as such for a significant period of time, at least one or more
sound crossing times (i.e., $\tau_{core} \geq 10^6$ yrs).

\subsubsection{The Source of the Confining Pressure}

The mean of the internal pressures of the core population is $<P/k> = 1.6 \times 10^5$ K
cm$^{-3}$.  As mentioned earlier this pressure corresponds to the average pressure within
the cores.  However the surface pressures of the cores are likely to be lower, since most
cores appear centrally concentrated and likely possess outwardly decreasing density
gradients.  We expect that the surface pressures of the cores are likely to be less by
factors of typically 2 than their individual mean pressures calculated here, i.e,
$<P/k>_{surface} \approx 8 \times 10^4$ K cm$^{-3}$.  The estimated mean surface
pressure of the cores is nearly an order of magnitude higher than the total (thermal +
turbulent) gas pressure of the interstellar medium (ISM), i.e.  $P_{\rm ISM}/k \approx
10^4$ K cm$^{-3}$ (Bertoldi and McKee 1992).  Thus the ISM is not likely the source of the
external pressure which confines the cores.  However, since the cores together represent
only about 1\% of the mass of the Pipe cloud, it is possible that the weight of the Pipe
cloud itself is the source of the external pressure within which the cores are embedded.
The pressure due to the weight of the Pipe cloud is given by:

\begin{equation}
 P_{\rm cloud} = (3\pi/20) G\Sigma^2\phi_G = 4.5 \times 10^3 \phi_G k A_V^2
\end{equation}
  
\noindent 
where $\Sigma$ is the mean mass surface density of the cloud i.e., $\Sigma = M_{cloud}/\pi
R^2$, k is Boltzmann's constant, $A_V$ is the corresponding mean extinction and $\phi_G$
is a dimensionless correction factor to account for the non-spherical geometry of the
cloud (Bertoldi \& McKee 1992).  Following the prescription of Bertoldi and McKee (1992)
we estimate $\phi_G$ to be 1.6.  The mean extinction we measure from the LAL06 data for
the Pipe cloud corresponds to $A_V \approx$ 4 magnitudes.  This yields $P_{\rm cloud}/k
\approx 10^5$ K cm$^{-3}$.  The close agreement between the estimated cloud pressure and
that of the cores indicates that the cloud, itself, is likely the source of the
external pressure for the cores.

We note here that turbulent ram pressure from the inter-core gas could also be a significant
source of external confining pressure for the cores.  The turbulent ram pressure is given by
$P_{\rm ram} = \rho \sigma^2$.  For the typical observed $^{13}$CO linewidths of 1 km
s$^{-1}$ (Onishi et al.  1999) and an assumed density for $^{13}$CO emitting gas of 10$^3$
cm$^{-3}$, we find $P_{\rm ram}/k \approx 5 \times 10^4$ K cm$^{-3}$, within a factor of 2
of the required pressure.  If turbulent pressure was the confining pressure for the cores
then the observed uniformity of the internal core pressures across the cloud would probably
require the turbulence to be driven on large, not small scales.  This is because the
downward turbulent cascade from large to small scales across the cloud would more likely
produce uniformity of turbulent motions on the scales of the cores than would turbulence
driven more locally (e.g., by outflows).  Indeed, if the pipe cloud is self-gravitating, it
is quite likely that the pressure due to the weight of the cloud is transmitted and manifest
by the turbulent velocity field.

Taken together the results presented in the preceding sections lead to a potentially
profound implication regarding the nature of the cores in the Pipe cloud.  The physical
structure of a dense core is dictated by a single requirement:  pressure equilibrium with
a surrounding source of external pressure.  The source of this external pressure is most
likely the weight of the molecular cloud in which the cores were formed.

\subsubsection{On the Origin of the Apparent Variations in Core Pressures}

One of the important characteristics of the observed relation between gas pressure and
core mass (figure \ref{totalpressure}) is the relatively large spread in the calculated
mean core pressures.  We briefly consider some possible causes of this spread.  One of
the most likely causes is a variation in the external pressure across the cloud.  The
cloud is certainly not uniform and variations of order a factor of 2 or so in the
pressure due to local variations in the weight of the cloud or the intercore 
turbulence in any one region could
certainly be possible.  In figure \ref{cloudpressure} we show the variation of core
pressure with position in the Pipe cloud.  It is clear that the cores in the "stem" of
the Pipe have systematically lower pressures and dispersion in pressures than cores in
the "bowl" of the Pipe.  Indeed, for cores at galactic longitudes $<$ 0 (stem) the mean
pressure is found to be $1.2 \pm 0.3 \times 10^5$ K cm$^{-3}$, whereas for cores in the
"bowl" the corresponding value is $1.9 \pm 1.2 \times 10^5$ K cm$^{-3}$.

Another plausible source of the variation in calculated core pressures is the presence
of {\sl static} magnetic fields that we did not account for in the calculation of the
internal core pressure.  We can estimate the magnitude of field strengths needed to
produce the pressure spread in figure \ref{totalpressure} by assuming that all the
cores are at exactly the same pressure.  We further assume that the entire variation
in the calculated pressures in figure \ref{totalpressure} is due to a variation in the
magnetic field strength within the cores.  The magnetic pressure, $P_B$, required to
bring the lowest pressure cores to the level of the constant external pressure is
then:  $P_B = P_{external} - P_{gas} = B^2/8\pi$, where $P_{gas}$ is the total
(thermal $+$ non-thermal) gas pressure and B the strength of the static field.  For
the values $P_{external}/k = 10^5$ and $P_{gas}/k = 2.5 \times 10^4$ K
cm$^{-3}$, we find B = 16 $\mu$G.  Thus, a variation in field strength between roughly
0-16 $\mu$G within the cores could produce the spread in pressures observed in figure
\ref{totalpressure}.  Static field strengths of order 16$\mu$G are perhaps smaller 
than might be expected but certainly reasonable for these densities (Crutcher 1999).
 
Another possible source for the variation in calculated pressures is that our
assumption of a constant gas temperature of 10 K in our calculation of gas pressure is
not strictly correct.  As mentioned earlier, \nht observations of a sample of
cores indicate that the actual gas temperatures vary between about 9.5 - 15 K
with a dispersion in the measured temperatures of $\sigma(T_K) = 2.3$ K
(Rathborne et al 2007).  However, this spread in temperature, if representative, could
induce a spread of only about 25\% in the actual pressures, too small to account
for the bulk of the observed spread in pressure.  It is also possible that our use of
\co lines to calculate the non-thermal pressure could introduce some spread in the
calculated pressures since, as mentioned earlier, the pressures derived from \co lines
are probably overestimates, because the \co lines simultaneously sample gas both in and
out of the cores.  This could conceivably induce a scatter of as much as a factor of
2 in some cores.  However, the pressures calculated using the NH$_3$ lines show a
spread similar in magnitude to those derived from the \co observations.  This is
another indication of the fact that the internal core pressures are dominated by
thermal motions.  Thus use of the CO lines to derive the non-thermal component of the
total pressure is not likely a source of the variation in the calculated gas
pressures.

In summary, the apparent variation in the internal core pressures is likely due to either
variations in the external pressure resulting from spatial variations in the structure of
the Pipe cloud, spatial variations in the intercore turbulence or to variations in the
amount of static magnetic field within the cores or to a combination of these effects.

\vskip 0.2in

\section{Implications and Discussion}

The close similarity in the shape of the Pipe CMF to that of the stellar IMF suggests a
one-to-one mapping of cores to stars, modified only by a constant star formation
efficiency (ALL06).  Thus these cores represent the final product of the cloud
fragmentation process (up to the present time).  If, for example, molecular clouds can be
characterized by a hierarchy of substructure produced by supersonic turbulence (Larson
1981), then these cores, with masses between 0.2-20 \msun\ and sizes of 0.1 - 0.4 pc must
define the physical scales for the termination of that process.  Because the vast majority
of these cores are starless the physical conditions that characterize them correspond to
the initial conditions for star formation.  In this paper we have combined previous
infrared extinction and molecular-line observations to determine the basic physical
properties (i.e., mass, size, density, internal pressure, etc.)  and to assess the nature
of this pristine population of pre-stellar cores.

We found the interesting result that the internal pressures of the cores are essentially
dominated by thermal motions across entire spectrum of core masses from 0.2-20 \msun.
Moreover, these pressures are of similar magnitude for all the cores, independent of their
mass, suggesting that the cores are in pressure equilibrium with their surroundings.  The
basic nature, structure and subsequent evolution of these cores is thus controlled by the
requirement of pressure equilibrium, self-gravity and the fundamental thermal physics
processes of heating and cooling.  For example, consider that heating by cosmic rays and
the interstellar radiation field coupled with cooling by molecular-lines keeps the cores
thermostated between 8-12 K, a small range of temperature (e.g., Goldsmith \& Langer
1978).  To maintain pressure equilibrium with their surroundings and support themselves
against gravity, the cores must adjust their overall density and density structure
appropriately.  It is thus not accidental that their mean densities span a small range.
Although most of the cores are gravitationally unbound, they are pressure confined and not
transient entities.  They likely evolve on acoustic time scales.  It is interesting in
this context that detailed molecular-line studies of the kinematics of two of the Pipe
cores, B68 and FeSt 1-457, suggest that these cores are oscillating around a state of
dynamical equilibrium and are thus likely to survive for at least a few sound crossing
times (Lada et al.  2003, Redman et al.  2004; Keto et al. 2006, Aguti et al.  2007).

Our observations provide important constraints for understanding the origin of core masses
and if there is a one-to-one mapping of the core to stellar mass, the origin of the
stellar IMF as well.  For example, one possible origin for the CMF is through
gravo-turbulent fragmentation.  On large scales the low density material in molecular
clouds is characterized by supersonic turbulence.  Indeed, the $^{13}$CO emission-lines
from the Pipe cloud are characterized by linewidths of $\approx$ 1 km s$^{-1}$ (Onishi et
al.  1999), representing bulk gas motions of about Mach 2.  In this context one important
constraint of our observations is the thermal and subsonic nature of the gas motions
within the cores.  This is in contrast to the results of numerical simulations of
turbulent fragmentation that typically produce dynamic cores characterized by supersonic
internal motions.  In one calculation Klessen et al.  (2005) found that under optimum
conditions only about 25\% of the cores produced by gravo-turbulent fragmentation would
appear to have subsonic turbulence to an observer.  This is inconsistent with our
observations of the Pipe cloud where $\sim$ 70\% of the cores are characterized by
subsonic turbulence (c.f.  figure \ref{ntsigmas}).  Another important constraint provided
by our observations is the fact that the cores are pressure confined entities and in
pressure equilibrium with an external pressure source, most likely provided by the weight
of the Pipe cloud itself.  As we discuss below, this constraint provides a potentially
critical insight into the origin of the core masses and the IMF.

\subsection{Core Stability: From CMF to IMF} \label{sec:CMF}

As thermally dominated, dense cores in pressure equilibrium with the surrounding cloud
material, the Pipe cores are perhaps most appropriately modeled as Bonnor-Ebert spheres
(e.g., Johnstone et al.  2000; Alves, Lada \& Lada 2001).  Bonnor-Ebert spheres are
pressure truncated isothermal spheres in hydrostatic and pressure equilibrium with their
surroundings.  Bonnor (1956) and Ebert (1955) investigated the stability of such
pressure-confined isothermal spheres and showed that under a specific condition such
objects became unstable.  This condition corresponds to the critical Bonnor-Ebert (BE)
mass given by:

\begin{equation}
m_{BE} = 1.82 \left({\overline{n} \over 10^4 cm^{-3}}\right)^{-0.5} \left({T \over 10 K}\right)^{1.5} 
\ M_\odot
\end{equation}

\noindent
where $\overline{n}$ is the mean volume density of the core.  Above this mass cores are
out of equilibrium and prone to fragmentation and/or collapse.  Below this mass cores are
in equilibrium states, primarily stable equilibrium states.  For the mean density of
cores in the Pipe cloud ($7.3 \times 10^3$ cm$^{-3}$), this critical mass is about 2
\msun.  In figure \ref{mBE} we plot the ratio of core mass to critical BE mass,
$m/m_{BE}$, (calculated individually for each core) against core mass.  The two quantities
form a tight relation which crosses the critical threshold ($m=m_{BE}$) at a mass of
$\approx 2-3$ \msun.  This is also the mass at which the cores appear to become
gravitationally bound (figure \ref{gbind}).

The results presented above may have interesting ramifications for understanding the
origin of the stellar IMF.  In our earlier study of the Pipe CMF we showed that its
overall shape was very similar to that of the stellar IMF for field stars and for the
young Trapezium cluster embedded in the Orion Nebula (ALL07).  However, the two functions
(i.e., CMF and IMF) differed in their characteristic masses.  The characteristic mass and
mass scale of the CMF was a factor of $\approx$ 3 higher than those of the stellar IMFs
which were very similar to each other.  This was interpreted to indicate that the stellar
IMF directly originates from the CMF after modification of the individual core masses in
the CMF by a uniform star formation efficiency (SFE) of $\approx$ 30\% (ALL07).
Theoretical investigations suggest that SFEs of this magnitude result from core disruption
via the outflows that are generated as a natural consequence of the formation and
evolution of accreting protostars (Matzner \& McKee 2000; Shu et al. 2004).  In this
picture each core must form one star or stellar system.

The observations in figure \ref{mBE}, however,  present a difficulty
for such an interpretation.  In order for the Pipe CMF to produce an IMF similar to that
of field stars or the Trapezium cluster, all the cores will have to form stars, yet as the
figure shows, a large fraction of the cores appear to be in stable configurations.  As it
stands now only cores with masses in the vicinity of, or greater than, the critical BE
mass will form stars.  If nothing else were to happen, the IMF of the stars that would
emerge from the cloud would, after adjusting for the SFE of $\sim$ 30\%, be similar to the
Salpeter IMF for stellar masses greater than about 0.6 - 1 \msun.  Lower mass stars would
also be expected to form as a result of random variations in the various important cloud
paramaters, such as cloud pressure, internal temperature, magnetic fields, etc.  (e.g.,
Adams \& Fatuzzo 1996).  But these stars would be rare compared to stars in the 0.6-1
\msun\ range.  Indeed, it would be extremely difficult to form any brown dwarfs.  It is
interesting that the IMF we just described is very similar to that derived by Luhman
(2004) for the Taurus cloud, a complex with similar overall properties to the Pipe.  The
Taurus IMF is unusual in that it does differ from the IMF of the field as well as the IMFs
of young clusters such as the Trapezium and IC 348 in that it has a peak near 1 \msun\
rather than at 0.1-0.3 \msun\ (Luhman 2004).

If the Pipe cores are to ultimately form an IMF more similar to the field star IMF, with
a peak closer to 0.1-0.3 \msun, then the lower mass cores must become unstable.  To
understand under which conditions this could happen we rewrite the critical BE mass in
terms of the external pressure, $P_{ext}$ and the sound speed, $a$:

\begin{equation}
m_{BE} = 1.15 \left({a \over 0.2 km s^{-1}} \right)^4 \left( {P_{ext}/k \over
10^5 K cm^{-3}}\right)^{-0.5}
\end{equation}

To stimulate the low mass cores to form stars we must lower the critical BE mass.  The
critical BE mass can be decreased in one of two possible ways.  First, we can increase
the external cloud pressure.  To lower the BE mass to that which would result in a
stellar IMF peak near 0.2 \msun, we need to decrease the BE critical mass to $\sim$ 0.7
\msun, about a factor of 3 lower.  If the sound speed, $a$, is kept constant we need to
increase the external pressure by about an order of magnitude to a value of $P_{ext}/k
\approx 10^6$ K cm$^{-3}$.  Given that the external pressure is provided by the weight of
the cloud, this would require an increase of about a factor of 3 in the cloud mass or a
decrease of just under a factor of two in the cloud radius (equation 2).  An increase in
the total cloud mass seems unlikely, but it is not inconceivable that the cloud could
gravitationally contract to nearly half its size if it somehow were to lose much of its
overall support against gravity, perhaps via the dissipation of its supersonic turbulence.

A second way to decrease the critical BE mass of the cores in the Pipe nebula would be to
decrease their internal pressures by decreasing the sound speed, $a$.  Because of the
sensitive dependence on $a$, the BE critical mass could be significantly decreased by a
small decrease in $a$.  To decrease the critical BE mass by the desired factor of 3 would
require only a 30\% decrease in the sound speed.  Since the sound speed depends on the
square root of temperature, a decrease in the kinetic temperature of about a factor of 1.7
could produce the desired decrease in critical mass.  For a 10 K core the temperature would
need to cool to about 6 K.  Since heating and cooling should thermostat the cores at $\sim$
10 K, cooling below this value would be somewhat difficult.  For starless cores the dust can
have equilibrium temperatures as low as 5-6 K so if there is any degree of gas-dus coupling
then the cores could be efficiently cooled by the dust to the required levels (Goldsmith
2001).  How likely this is given the typical densities of the Pipe cores is difficult to
assess.  Loss of whatever small amount of magnetic or (subsonic) turbulent pressure the
cores possess could also decrease the internal pressure and facilitate their eventual
collapse.  Ambipolar diffusion could naturally lead to the desired decrease in the internal
magentic pressure (e.g., Adams \& Shu 2007).  A combination of some cooling and loss of
magnetic and turbulent pressure support is quite possible and eventually many of these low
mass cores could reach the critical mass threshold.

One potential difficulty with the above evolutionary scenario would arise if the timescale
for the presently stable cores to evolve to unstable configurations is greater than a few
million years.  Since the more massive unstable cores will form stars relatively quickly,
the end result of the star formation process would be a stellar population displaying a mass
dependent age gradient.  Although age spreads in young stellar populations can be on the
order of 3- 5 Myr, no evidence has yet been found for a systematic, mass dependent age
gradient of similar magnitude.  

It also may be difficult to increase the external pressure
provided by the Pipe cloud to the extent necessary to push the lowest mass cores to
collapse, form stars, and thereby produce a fully sampled, standard IMF.  However, such high
pressures do appear to characterize other star forming regions, particularly those where
clusters and thus most stars are formed (Lada \& Lada 2003).  For example, Johnstone et al.
(2000, 2001) have estimated $P/k \sim 10^{7} \ \rm{and} \ 10^6$ K cm$^{-3}$ for the $\rho$
Oph and Orion B cluster-forming clouds, respectively.  Moreover, consider the cores of
Infrared Dark Clouds (IRDCs).  These opaque, massive regions are believed to be the
precursors of embedded clusters (Rathborne, Simon \& Jackson 2006).  A survey of IRDCs for
millimeter-wave emission from dust provides mass and size estimates for 140 cold dense cores
within IRDCs (Rathborne et al.  2006).  The median mass and radius for this sample of cores
are 121 \msun\ and 0.27 pc, respectively.  The corresponding pressure from equation 2 is
$P_{ext}/k = 2.7 \times 10^6$ K cm$^{-3}$.  For such levels of pressure the corresponding
critical BE mass could decrease substantially.  However, this is only true if the sound
speed, $a$, remained similar to that in the Pipe cloud.  It is more likely that in such
compact massive regions the sound speed in equation 4 is higher than that of a 10 K purely
thermal gas.  Given the sensitivity of the BE critical mass to $a$, this could easily
compensate for the increase in external pressure, leaving the critical BE mass similar to
that in the Pipe.  Indeed, this appears to be the case for the cores in the Ophiuchus and
Orion B clouds studied by Johnstone et al.  (2000, 2001).  For the Ophiuchus cores the
typical kinetic temperature was estimated to be $\sim$ 20 K and the corrresponding surface
pressure $P/k \sim 3 \times 10^6$, comparable to the external pressure exerted by the cloud
itself, similar to the situation for the Pipe.  The critical BE mass in that cloud was found
to be $\sim$ 1-2 \msun\ with most of the cores being subcritical, again similar to the
situation in the Pipe.  For the cores in Orion B the typical kinetic temperature was found
to be $\sim$ 30 K with the corresponding surface pressure $\sim 10^6$ K cm$^{-3}$ and the
critical BE mass $\sim$ 3 \msun.  However, given the larger values for $a$ and internal
temperatures in such high pressure regions, cores with masses below the critical BE mass may
have more room to cool and lose internal pressure support than similar cores in lower
pressure regions like the Pipe.  Thus it may be easier to produce low mass and even
substellar objects in clustered environments than in regions similar to the Pipe or Taurus
clouds.  Because the formation of brown dwarfs does require restrictive conditions, one
might expect the substellar portion of the IMF in a recently formed stellar population to be
very sensitive to local conditions and thus to vary noticeably from region to region.

In the preceeding discussion of the possible evolution of the cores in the Pipe CMF, it was
assumed that the core masses were fixed and unchanging.  Another possibility worthy of
consideration is that the core population in the Pipe is so young that the cores have not
yet obtained their final masses.  This, in particular, could have important consequences for
the evolution of the presently stable low mass cores in the cloud.  Separate studies by
Clark and Bonnell (2005) and Gomez et al.  (2007) have raised the possibility that dense
cores formed from a turbulent medium could grow in mass with time, starting out as unbound
low mass objects and ending up as bound objects at the threshold of collapse and
fragmentation.  In this picture the stable but unbound cores in the Pipe would be still
gaining mass and as a result will eventually cross the critical BE threshold and ultimately
collapse to form stars.  The resulting IMF would not bear any relation to the original CMF.
Whether such models can account for the basic physical properties of the cores reported here
remains to be determined.

\subsection{The Origin of the CMF: Thermal Fragmentation Under Pressure}
\label{sec:IMF}

The fact that the critical BE threshold for the entire core population corresponds to a
mass of $\sim$ 2 \msun, as shown by figure \ref{mBE}, is potentially very significant.
This becomes apparent when one considers the core mass function (CMF) of the Pipe cloud
derived from the extinction observations by ALL07.  These authors found the CMF to rise,
with decreasing mass, in a power-law fashion, from the highest mass ($\sim$ 20 \msun) core
to roughly 2-3 \msun (similar to the behavior of the Salpeter or stellar IMF).  At this mass
the CMF was found to break or depart from the power-law.  The CMF then continued to rise
only very slowly, forming a broad peak between 2 - 0.4 \msun, before declining toward
lower masses (c.f.  figure 3 of ALL07).  The CMF indicates that most of the cores that
formed in the Pipe cloud formed with masses between 0.4 - 2.0 \msun.  The departure point
from the Salpeter-like power-law sets a characteristic mass of $\sim$ 2-3 \msun\ for the CMF
in this cloud.  The fact that this characteristic mass is very close to the critical BE
mass for the cores is unlikely to be a chance coincidence, indeed it provides a
potentially interesting clue concerning the very origin of the CMF.  The physical
interpretation of this result is straightforward:  the characteristic mass of the CMF (and
the CMF itself) is the direct result of thermal fragmentation in a pressurized medium.  In
other words, the CMF may have its origin in the physical process of pressurized thermal
fragmentation.

It is instructive in this context to express the core mass function, $CMF(logm)$, in terms
of the characteristic mass, $m_c$, a shape parameter, $s_i$ and an arbitrary constant,
$c_0$, as follows:

\begin{equation}
CMF(logm) = c_0 \Psi(log(m/m_c), s_i)
\end{equation}

\noindent 
The results of this paper suggest that $m_c = m_{BE}$.  
If the functional form of the stellar IMF is the same as that of the CMF but with a
characteristic mass reduced by the SFE then we can
express the IMF as follows:

\begin{equation}
IMF(logm) = c_1 \Psi(log(m/m_c^*), s_i)
\end{equation}

\noindent
with,

 $$m_c^* = m_{BE} \cdot SFE$$

Expressed in this way these two equations suggest a possible generalization of our
results.  If we assume that the shape of the CMF is invariant, that is, $s_i$ is constant,
then the CMF is completely specified by one parameter, the critical BE mass, which in turn
depends on only two simple physical parameters, external pressure, $P_{ext}$, and sound
speed, $a$ (i.e., equation 4).  This has the powerful implication that knowledge of
the pressure and sound speed could be used to predict the CMF (and ultimately the IMF)
that would be produced in any given star forming environment, near or far, past or
present.  In principle, both parameters $a$ and $P_{ext}$, can be directly obtained or
inferred from observations.

Whether our conjecture of an invariant shape for the CMF is a realistic one however
remains to be verified by both observation and theory.  This conjecture can be directly
tested observationally with additional extinction and dust emission surveys of dense gas
in other molecular cloud complexes.  Indeed, numerous recent determinations of the CMFs
using observations of dust emission from cores in a number of other clouds certainly
appear to support our conjecture (e.g., Motte et al.  2001; Testi \& Sargent 1998;
Johnstone et al.  2000, 2001; Beuther \& Schilke 2004; Enoch et al.  2006; Stanke et al.
2006; Walsh et al.  2006).  Moreover, if the stellar IMF originates directly from a
one-to-one transformation of the CMF, then the observed similarity of the IMFs in young
clusters with that of the field perhaps also points to an invariant CMF shape function.
In addition theoretical considerations have suggested that the CMF (and IMF) could be
characterized by a simple log-normal form or shape.  This is a natural outcome of the
central limit theorem if a number of independent physical variables contribute to the
final determination of initial core masses (e.g.  Adams \& Fatuzzo 1996).  On the one
hand, this has to some extent been borne out by numerical simulations of gravitational
fragmentation (e.g., Klessen, Burkert \& Bate, 1998).  On the other hand, simulations of
turbulent fragmentation do not produce a universal CMF (Ballesteros-Paredes et al.  2006).
Although recent observations support the conjecture of a universal shape for the CMF, the
theoretical situation is unclear.

%

\subsection{Speculations on the Origin of Stellar Multiplicity} \label{sec:binarity}

Stellar multiplicity is a fundamental parameter of stellar systems.  It is an increasing
function of stellar mass.  The single star fraction (SSF) for M stars (m $\sim$ 0.1-0.6
\msun) is measured to be around 70\% and since the vast majority of stars formed are M
stars, most stars produced in the Galaxy are single (Lada 2006).  The SSF declines
steadily along the Salpeter, power-law portion of the IMF with the most massive (OB) stars
being characterized by a SSF $\leq$ 20\%, (i.e., corresponding to the highest multiplicity
fraction).  Existing theoretical attempts to account for stellar multiplicity statistics
(e.g., Durisen et al.  2001; Kroupa 1995) are unable to simultaneously explain both the
mass dependence and the overall magnitude of the SSF.  For the Pipe cloud the peak of the
CMF nearly coincides with the critical BE mass.  It is difficult to imagine how critical
mass, thermally supported, hydrostatic cores in pressure equilibrium with their
surroundings could fragment any further to form binary or multiple star systems.  Such
cores are more likely to directly collapse in an inside-out fashion to form a single star
(Shu 1977; Shu, Adams \& Lizano, 1987; Foster \& Chevalier 1993).  However, cores whose
masses exceed the critical mass are not only highly unstable but also out of equilibrium,
having masses ranging from just above $m_{BE}$ to $\approx$ 10\ $m_{BE}$ (figure
\ref{mBE}).  Numerical simulations of the collapse and fragmentation of isothermal clouds
whose masses exceed the Jeans mass suggest that the number of fragments formed in such a
process is comparable to the initial number of Jeans masses in the cloud with the result
that binary systems and hierarchical multiple systems are frequently produced (Larson 1978;
Boss \& Bodenheimer 1979).  Thus cores with masses in excess of the critical BE mass
should be increasingly susceptible (with increasing mass) to fragmentation and multiple
star formation.  The outcome of star formation in these cores might thus be expected to
produce an increasing multiplicity fraction with mass along the Salpeter, power-law
portion of the IMF, similar to what is observed.  In this context it also is interesting
to note that B59, the most massive core in the Pipe (and the only one so far known to form
stars), has a measured mass of $\sim$ 20 \msun\ ($\sim$ 10 $m_{BE}$) and has already
fragmented and produced a group of $\sim$ 13 low mass young stars (Brooke et al 2006)
\footnote{
The production of increasing multiplicity with core mass along the Salpeter portion
of the CMF suggests that a departure from a strict one-to-one mapping of the CMF to
the IMF may occur for the more massive cores.  The overall similarity of
the CMF to the IMF over most of this regime would seem to indicate that cores in the
2-10 \msun\ range produce binary and multiple systems in which the primary stars
dominate the masses of the stellar systems. However, for the
most massive cores (e.g., B59) that produce stellar clusters, the concept
of a strict one-to-one mapping between the two mass functions is not
likely to remain valid.}.
For the most part cores with masses less than the critical mass are presently stable
against collapse.  However, if these cores experience either a slow increase of external
pressure or a gradual decrease in internal pressure or some combination of both effects,
they will cross the equilibrium threshold at just the critical BE mass.  Thus they are
also likely to produce single stars.  Since these objects and the presently critically
stable objects represent the peak of the CMF and the bulk of the cores formed, we expect
most stars formed in the Pipe to be single, similar to what is observed for field
stars (Lada 2006).  If the CMF of the Pipe is the result of thermal fragmentation in a
pressurized medium, and is transformed directly into the IMF by the SFE, then the overall
magnitude and mass dependence of stellar multiplicity may be a natural outcome of this
straightforward core fragmentation process.

\section{Summary and Conclusions} 

We have combined previous infrared extinction and millimeter-wave molecular line
observations to determine the physical nature of the population of dense cores
in a single molecular cloud, the Pipe Nebula.  We summarize the primary results of the
paper as follows:

\noindent
1)\ The cores are found to be relatively dense objects that display a narrow range in
 number density with a median value of $n = 7.1 \pm 2.1 \times 10^3$ cm$^{-3}$.

\noindent 
2)\ Widths of C$^{18}$O and NH$_3$ lines observed toward the cores are not
correlated with core size or mass and do not obey a linewidth-size relation.  The
non-thermal velocity dispersions measured in both tracers are also independent of core
size and mass and are predominately characterized by subsonic (70\%) or trans-sonic (25\%)
motions.

\noindent
3)\ The ratio, $R_p$, of thermal to non-thermal gas pressure is found to range between
roughly 0.2 - 100 and thermal pressure is found to exceed non-thermal (turbulent) gas
pressure in the large majority ($\sim$ 67-80\%) of the cores.  Thermal pressure support is
significant (i.e., $R_p > 0.5$) for nearly all ($\sim$ 90\%) the cores.

\noindent 
4)\ The core internal pressures are surprisingly similar over the entire 0.2-20
\msun\ range of core mass and exceed the expected total gas pressure of the ISM by nearly
an order of magnitude.  The similarity of their internal pressures indicates that the
cores are in pressure equilibrium with an external pressure source.  The source of this
external pressure is likely provided by the overall weight of the Pipe cloud in which the
cores are embedded.

\noindent 
5)\ The dispersion in internal core pressure of about a factor of 2-3 is
significant and likely results from either local variations in the external pressure due
to structural variations in the Pipe cloud or the presence of internal static magnetic
fields with strengths between 0--16 $\mu$G or a combination of both.

\noindent
6)\ Only the most massive ($m \geq 2$ \msun) cores are gravitationally bound. Although
the majority of cores are gravitationally unbound they appear to be pressure-confined,
coherent objects.

\noindent
7)\ The entire core population is found to be characterized by the same critical
Bonnor-Ebert mass of $\sim$ 2 \msun.  This mass is very similar to the observed
characteristic mass of the CMF in this cloud (ALL07).  This in turn may suggest that the
CMF, the direct progenitor of the stellar IMF, originates as a result of thermal
fragmentation in a pressurized medium.

This last conclusion is potentially very significant because it suggests that the CMF that
is produced out of a molecular cloud may be able to be specified by only a few basic
physical parameters, such as external pressure and temperature, to name two.  Moreover,
the structure and evolution of dense cores may depend on the interplay of only a small and
restricted set of basic physical parameters/processes such as self-gravity, heating and
cooling of the core gas and the pressure of the external medium.  

Another key finding of our study is that the core formation process simultaneously
produced many objects that are in apparently stable configurations in addition to a number
of bound, non-equilibrium objects.  The presently stable cores are unlikely to undergo
star formation unless they experience further evolution driven by either an increase in
external pressure or a decrease in their internal pressures or a combination of both.  If
only those cores that are presently bound, and either out of equilibrium or critically
stable, collapse to form stars, the stellar IMF that will emerge from this cloud will
resemble that of the Taurus star forming region, not that of the field or of embedded
young clusters where most stars are formed.  It would also appear very difficult to form
substellar objects from a set of such pressure confined cores.  The formation of lower
mass stars, brown dwarfs and a standard IMF may be facilitated in regions characterized by
higher external pressures, such as massive cluster forming cores.

Since most cores produced in a pressurized thermal fragmentation process appear to have
masses near the critically stable value, the stars that form from them would tend to be
single.  The observed increase in stellar multiplicity with mass along the Salpeter
portion of the stellar IMF may be a natural consequence of the additional production of
cores in this type of fragmentation process that have masses increasingly in excess of the
critical value and thus increasingly likely to form multiple stellar systems.

Finally, the most important result of our study may be the recognition of the significant
role played by pressure in determining both the physical natures of dense cores and the
process that leads to their formation and as a result to the development of the dense core
and initial stellar mass functions.

\vskip -0.1in

\acknowledgments

We indebted to Frank Shu, Ramesh Narayan, and Doug Johnstone for
enlightening discussions. We thank the referee Ian Bonnell for
criticisms and suggestions that improved the paper.
This research was supported in part by NASA Origins grant
NAG-13041.

\section{APPENDIX}

The basic physical parameters of the Pipe extinction cores calculated from the
infrared extinction observations of Lombardi et al. (2005) are listed in the
table. These respectively include the core identification number (ID), mass,
radius and density. The mass is the background-subtracted mass derived from the 
wavelet decomposition of the infrared extinction map by Alves et al. (2007) assuming
a standard gas-to-dust abundance.



\begin{figure}
\begin{center}
\includegraphics[height=5in]{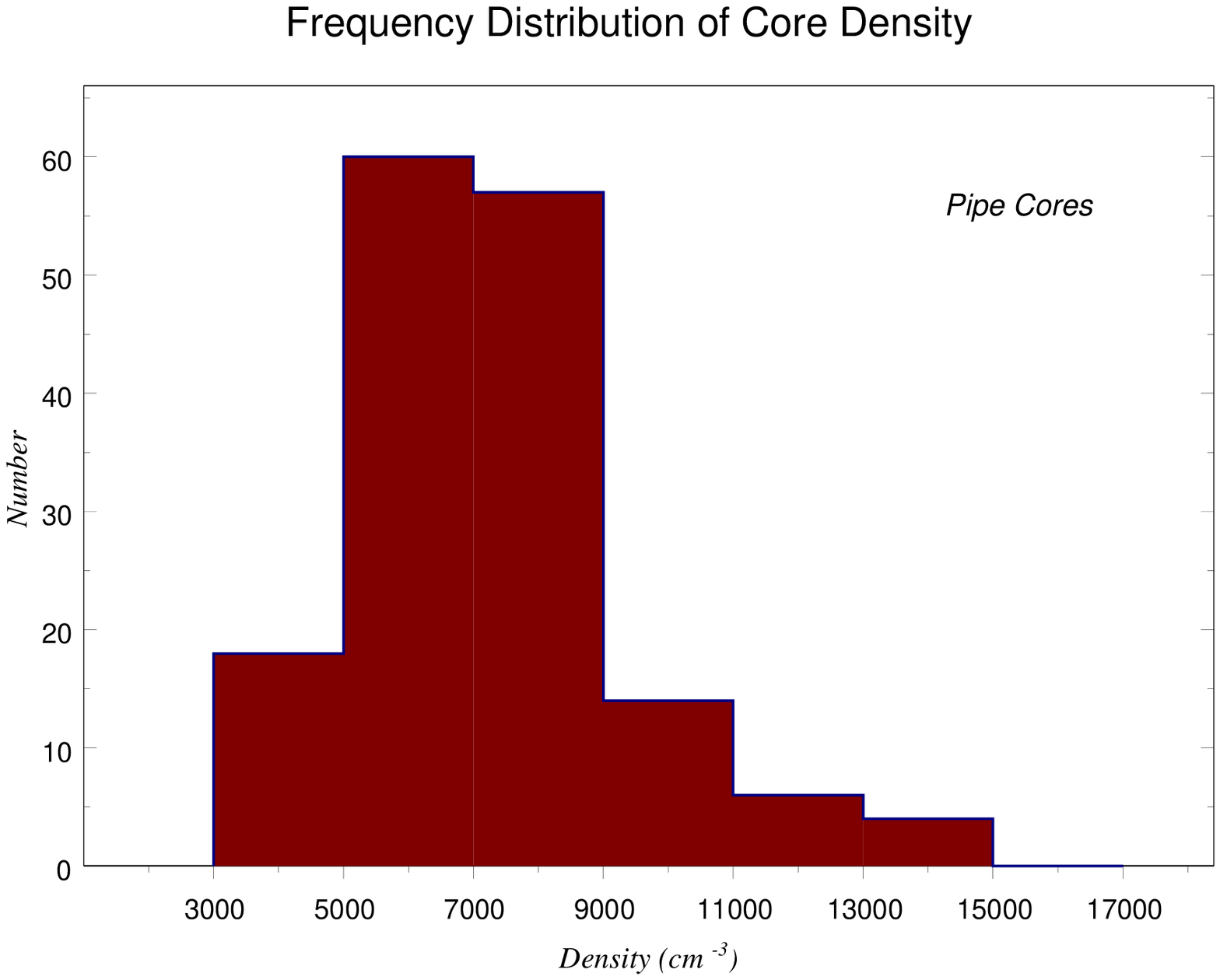}
\caption{The frequency distribution of mean core particle densities (cm$^{-3}$). The
extinction cores identified in the Pipe cloud are dense cores with a relatively
narrow spread in mean density}
\label{coredensity}
\end{center}
\end{figure}

\clearpage

\begin{figure}
\begin{center}
\includegraphics[height=5in]{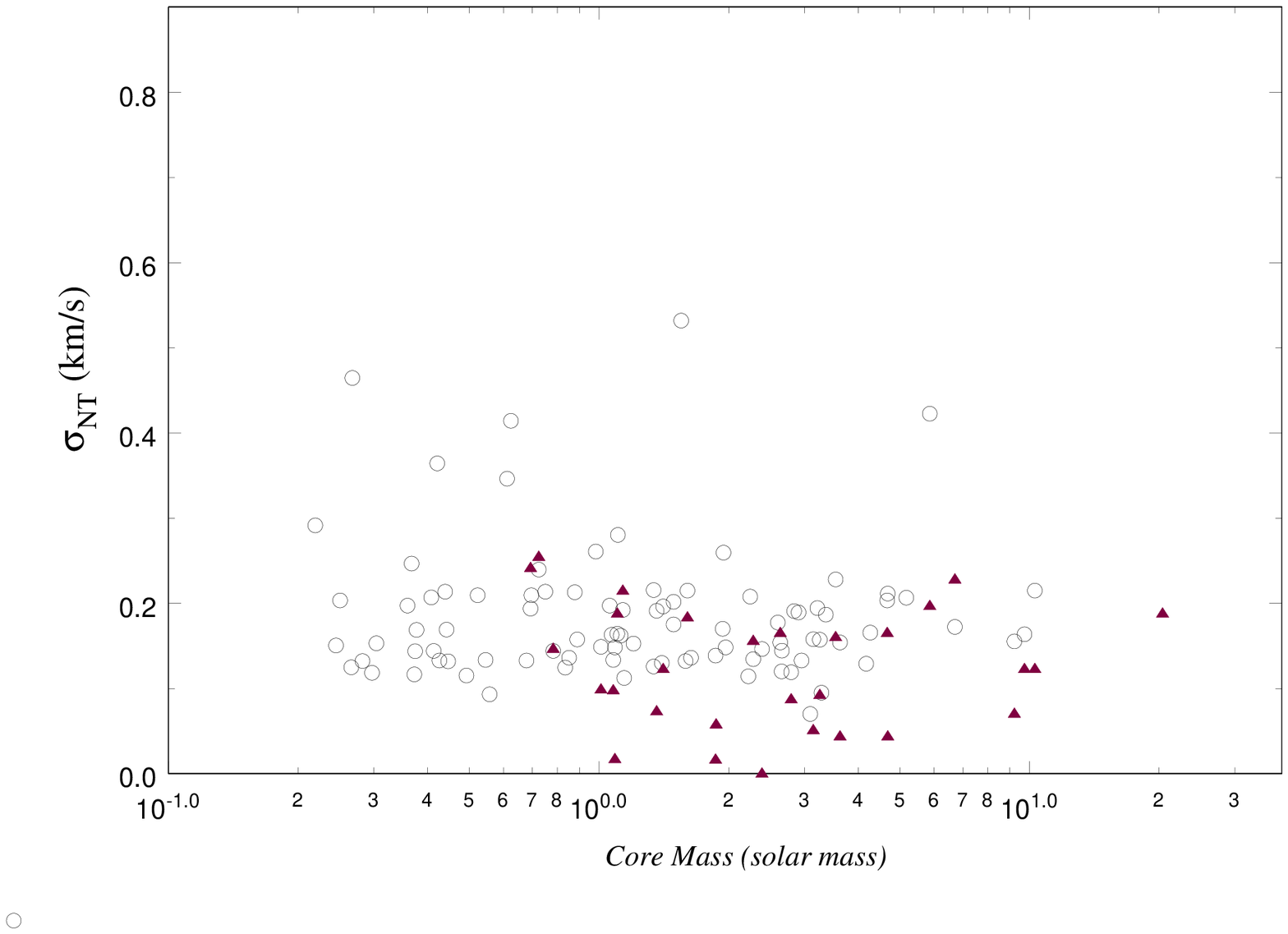}
\caption{The relation between non-thermal velocity dispersion and mass for
the dense core population of the Pipe Nebula. These dispersions are predominately
subsonic and not correlated with core mass. Open symbols (circles) correspond to
 \co measurements and the filled symbols (triangles) to \nht measurements}
\label{ntsigmas}
\end{center}
\end{figure}

\clearpage

\begin{figure}
\begin{center}
\includegraphics[height=5in]{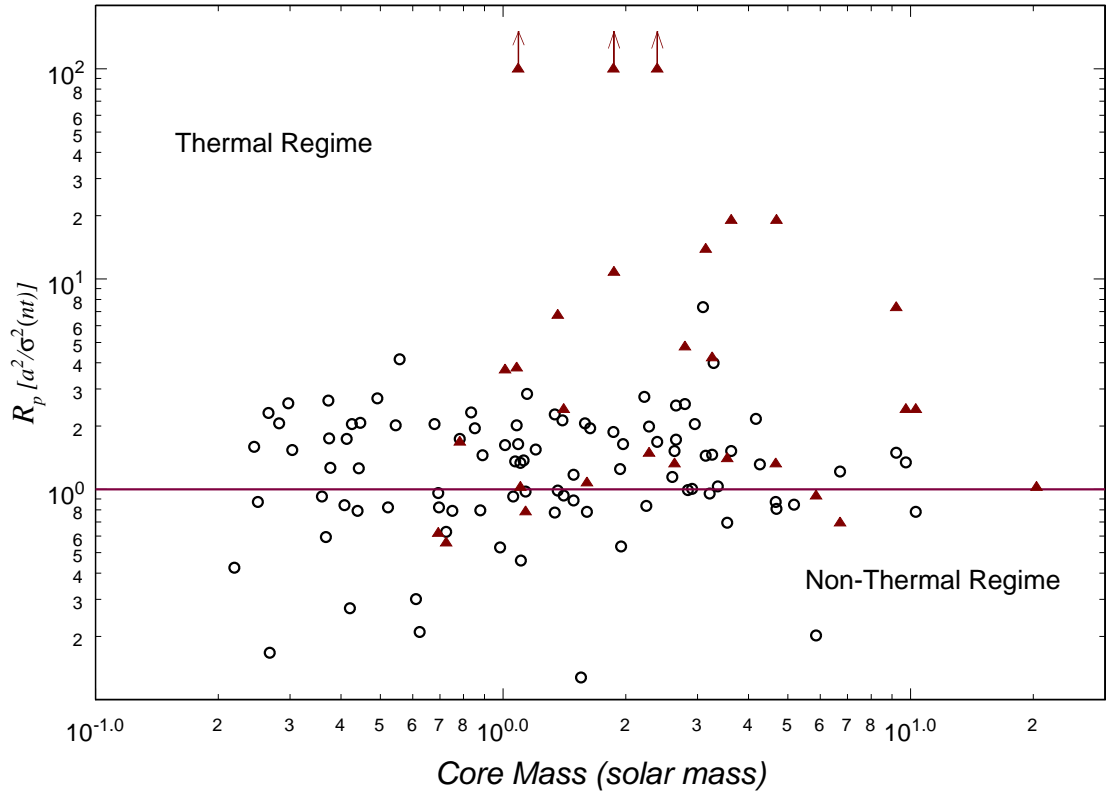}
\caption{The ratios of thermal to non-thermal gas pressure in the Pipe cores.
Thermal pressure dominates the internal pressure for the great majority of the cores.
Thermal pressure is a significant pressure source ($R_p > 0.5$) for nearly all
($\sim$ 90\%) the core population. The three lower limits correspond to 
three cores whose \nht linewidths were indistinguishable from purely 
thermal profiles in a 10 K gas. Symbol key identical to Figure \ref{ntsigmas}.}
\label{Rpressures} 
\end{center}
\end{figure}

\clearpage

\begin{figure}
\begin{center}
\includegraphics[height=5in]{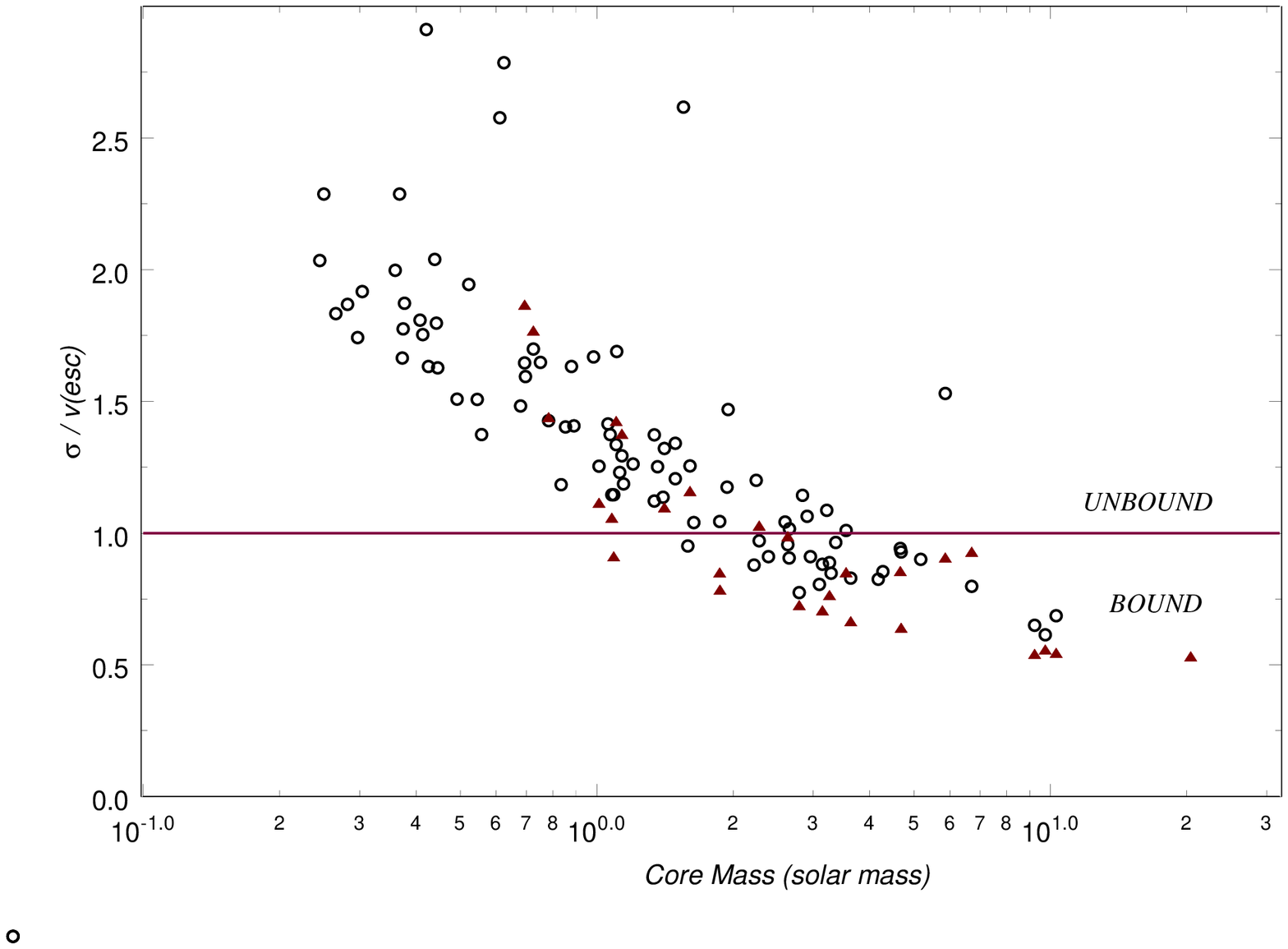}
\caption{The ratio of total (3D) velocity dispersion to escape velocity plotted 
against core mass. Most cores in the Pipe cloud appear to be gravitationally
unbound. Symbol key identical to Figure \ref{ntsigmas}.}
\label{gbind}
\end{center}
\end{figure}

\clearpage

\begin{figure}
\begin{center}
\includegraphics[height=5in]{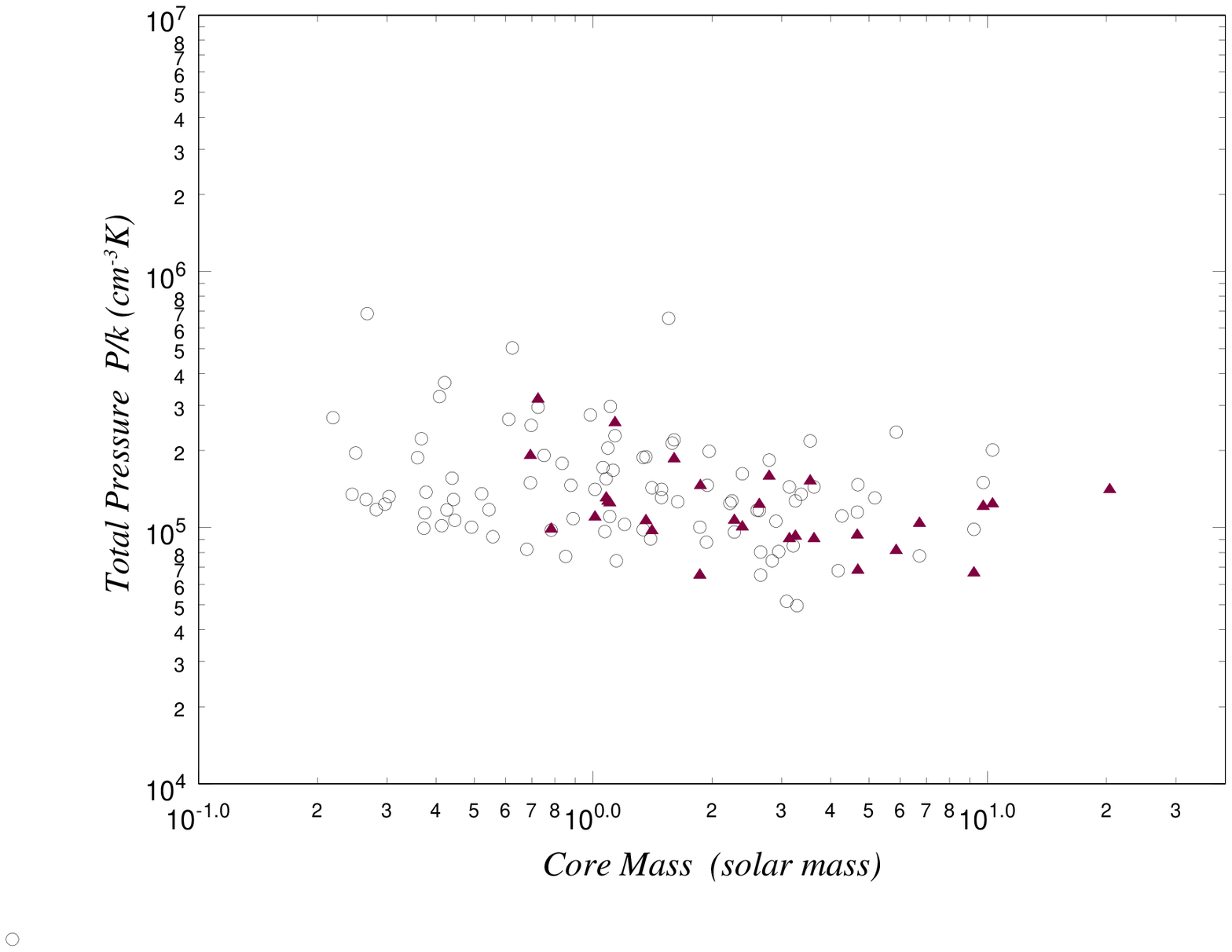}
\caption{Total internal gas pressure plotted as a function
of core mass. The close similarity of core pressures across the
entire span of core mass suggests that the individual cores are
in a state of pressure equilibrium with an external pressure source.
Otherwise symbols same as in Figure \ref{ntsigmas}.  }
\label{totalpressure}
\end{center}
\end{figure}

\clearpage

\begin{figure}
\begin{center}
\includegraphics[height=5in]{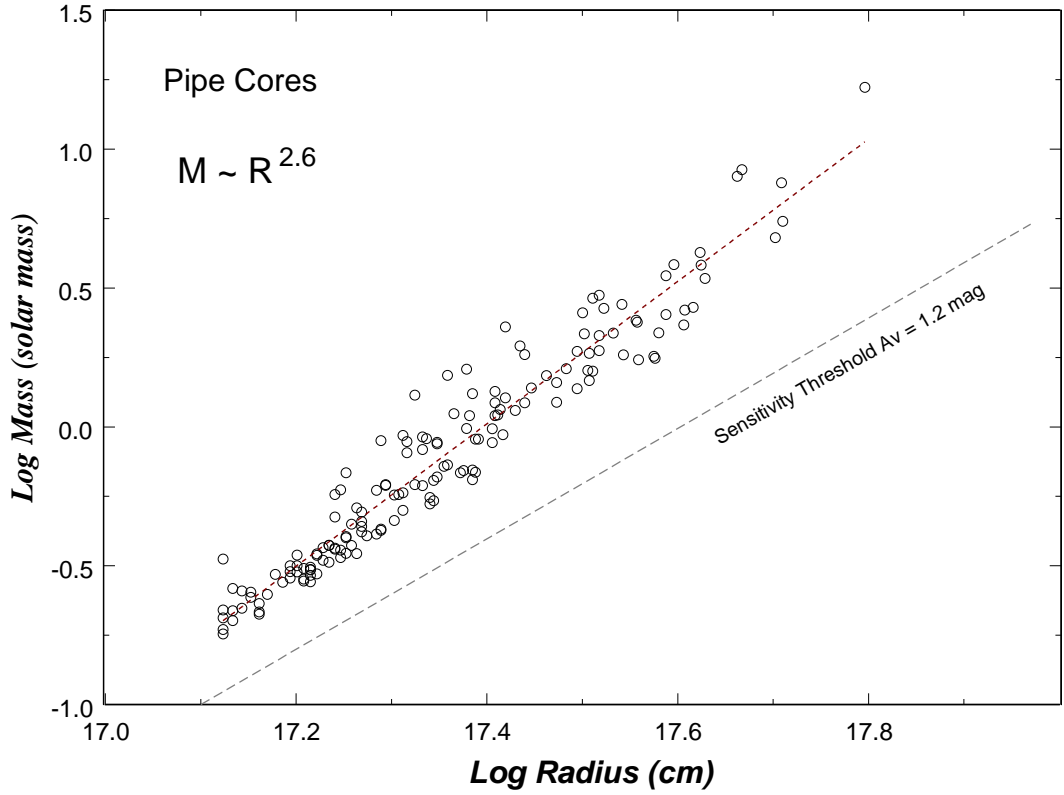}
\caption{The mass-radius relation for the dense cores in the Pipe Nebula.
The short-dashed line represents the least-squares fit to the data. The long-dashed
line indicates the sensitivity threshold of the observationns.}
\label{mvsr}
\end{center}
\end{figure}

\clearpage

\begin{figure}
\begin{center}
\includegraphics[height=5in]{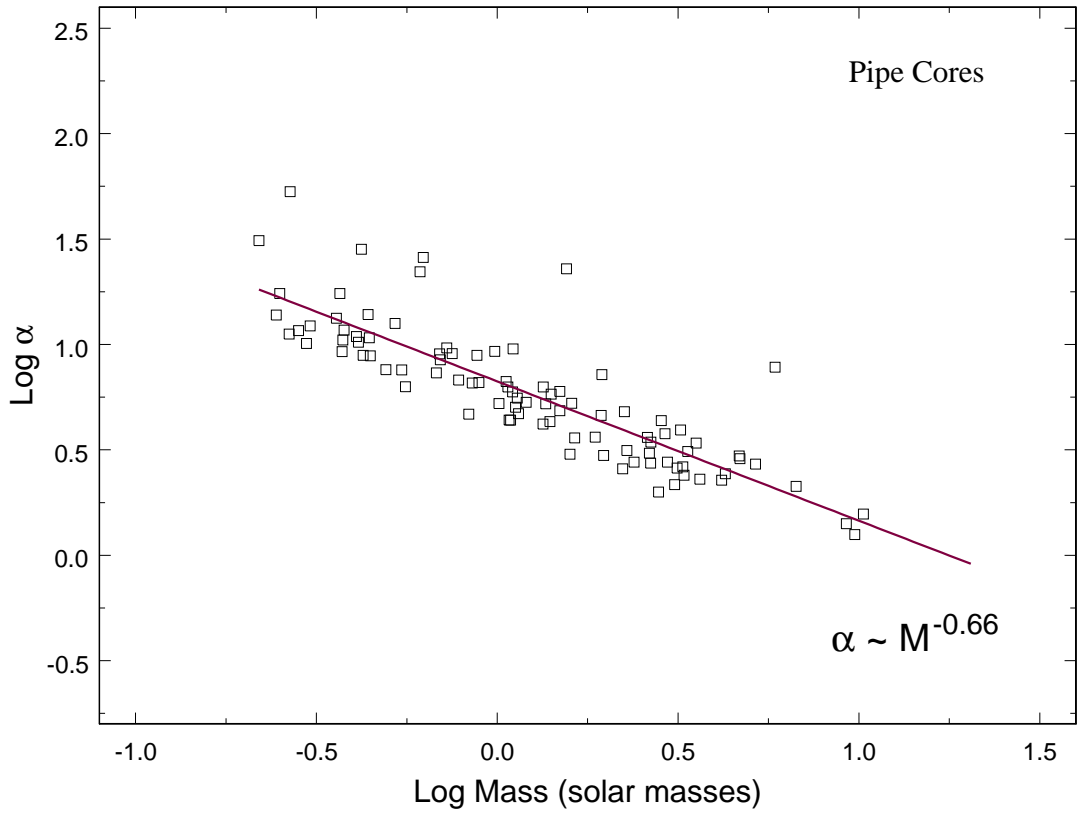}
\caption{The relation between the virial parameter and core mass.
The solid line is the least-squares fit to the data. }
\label{virialparameter}
\end{center}
\end{figure}

\clearpage

\begin{figure}
\begin{center}
\includegraphics[height=5in]{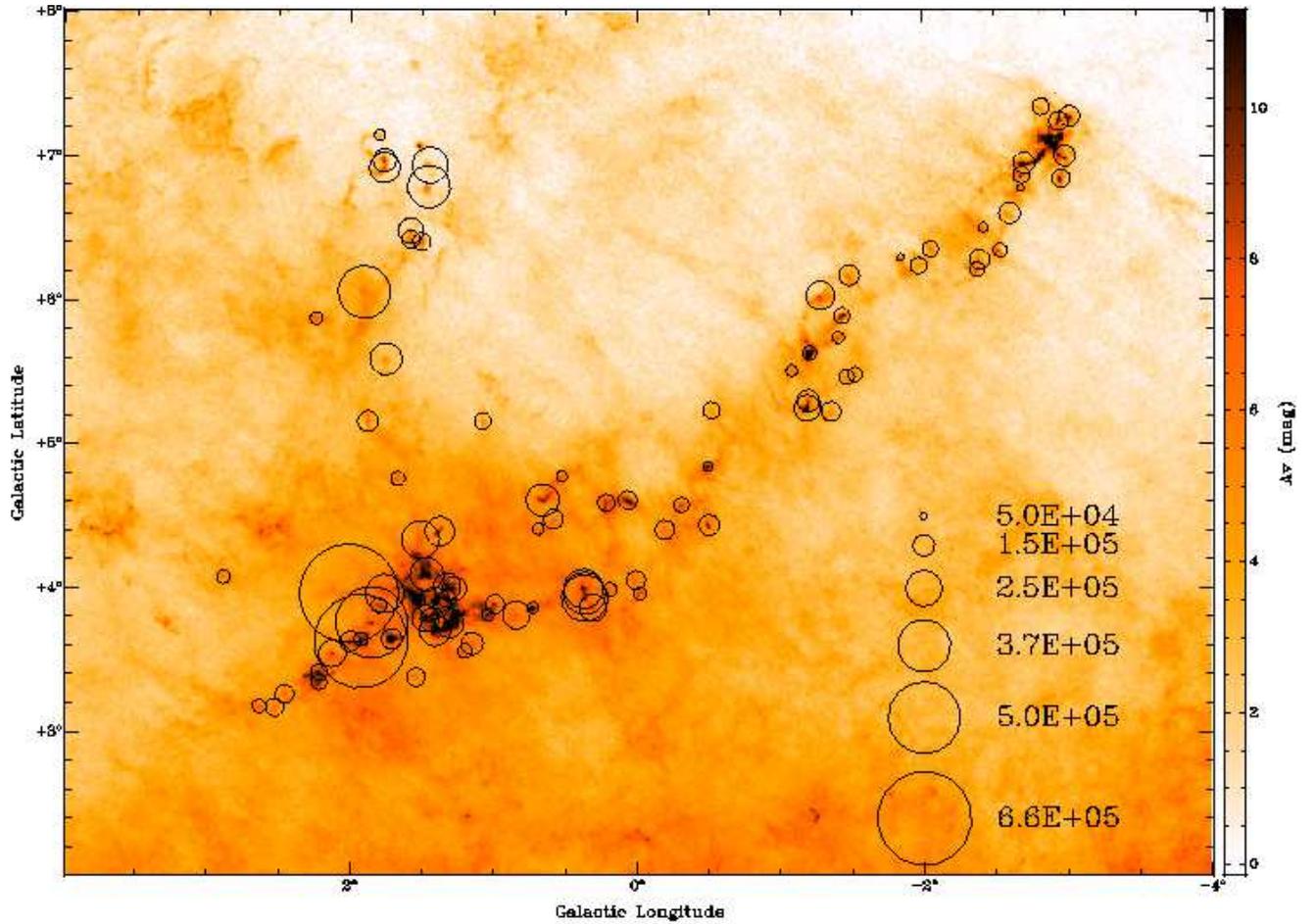}
\caption{The spatial variation of internal core pressures across 
the Pipe cloud. The core pressures are represented by open circles 
whose size (area) is proportional to the total internal gas pressure
in each core. These pressures are plotted on top of the near-infrared
wide field extinction map of Lombardi, Alves \& Lada (2006).}
\label{cloudpressure}
\end{center}
\end{figure}

\clearpage

\begin{figure}
\begin{center}
\includegraphics[height=5in]{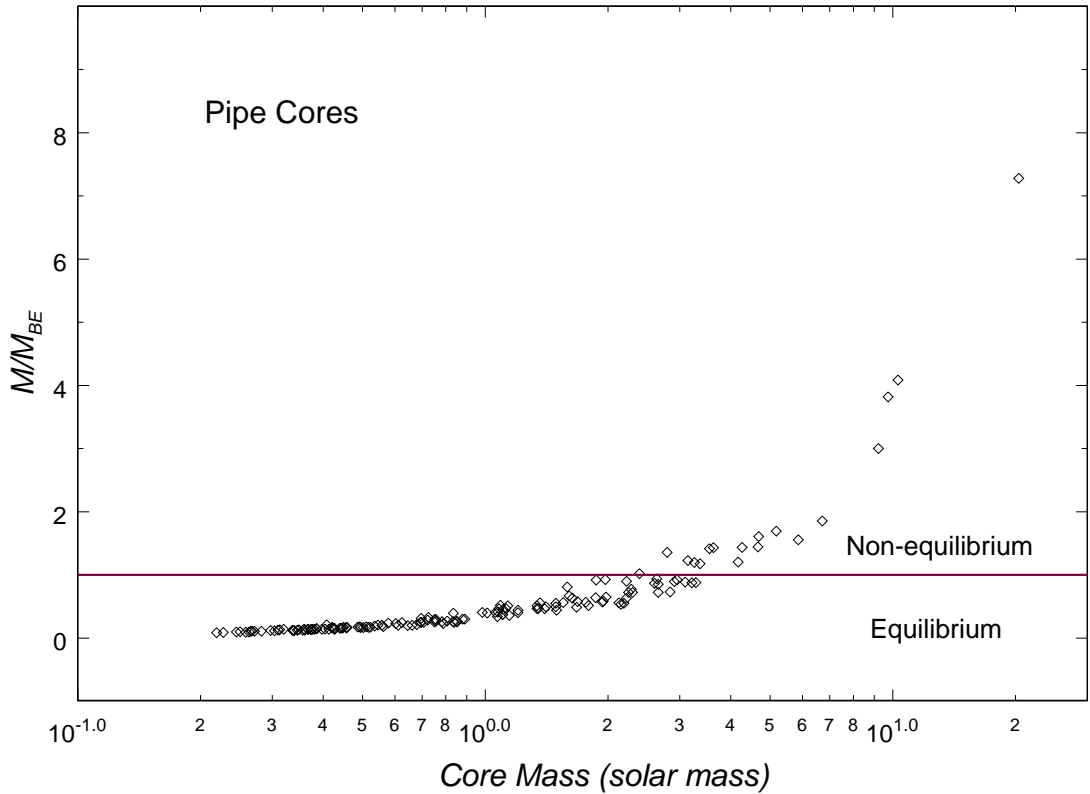}
\caption{The ratio of core mass to Bonnor-Ebert critical mass
for each individual core plotted against core mass. The entire core population appears to
be characterized by a single critical BE mass of $\approx$ 2 \msun. Cores with masses
in excess of the critical mass are likely out of equilibrium and destined to form
stars. There is also a large population of cores that are presently in equilibrium
states. Most of these are likely in stable equilibrium states and thus are unlikely
to collapse to form stars unless further perturbed via an increase in the external
pressure, loss of internal pressure support (e.g., cooling), or a combination of 
both effects. }
\label{mBE}
\end{center}
\end{figure}

\clearpage


\begin{deluxetable}{rrrr}
\tablewidth{0pt}
\tablecaption{Physical Core Properties \label{tab:data}}
\tablehead{
\colhead{Pipe Core} &
\colhead{Mass} &
\colhead{Radius} &
\colhead{Density}  \\
\colhead{(ID)} &
\colhead{\msun} &
\colhead{cm (10$^{17}$)} &
\colhead{cm$^{-3}$ (10$^4$)}  
}
\startdata
    1  &    0.38  &    1.85  &    0.73 \\
    2  &    0.46  &    2.04  &    0.66 \\
    3  &    0.49  &    2.12  &    0.63 \\
    4  &    0.36  &    1.85  &    0.69 \\
    5  &    0.23  &    1.50  &    0.82 \\
    6  &    3.14  &    3.57  &    0.84 \\
    7  &    4.69  &    4.45  &    0.65 \\
    8  &    3.26  &    3.76  &    0.75 \\
    9  &    0.56  &    2.27  &    0.59 \\
   10  &    0.51  &    2.09  &    0.68 \\
   11  &    3.37  &    3.93  &    0.68 \\
   12  &   20.37  &    7.06  &    0.71 \\
   13  &    0.54  &    2.04  &    0.78 \\
   14  &    9.73  &    5.19  &    0.85 \\
   15  &    2.64  &    3.58  &    0.70 \\
   16  &    3.29  &    4.66  &    0.40 \\
   17  &    0.69  &    2.27  &    0.72 \\
   18  &    0.35  &    1.82  &    0.70 \\
   19  &    0.34  &    1.73  &    0.79 \\
   20  &    2.28  &    3.53  &    0.64 \\
   21  &    2.66  &    4.29  &    0.41 \\
   22  &    1.01  &    2.42  &    0.86 \\
   23  &    1.87  &    3.27  &    0.65 \\
   24  &    0.70  &    2.29  &    0.71 \\
   25  &    1.10  &    2.78  &    0.63 \\
   26  &    0.37  &    1.85  &    0.72 \\
   27  &    3.09  &    4.37  &    0.45 \\
   28  &    0.32  &    1.54  &    1.07 \\
   29  &    0.43  &    2.07  &    0.59 \\
   30  &    0.41  &    1.99  &    0.64 \\
   31  &    1.95  &    3.61  &    0.50 \\
   32  &    0.45  &    1.97  &    0.72 \\
   33  &    4.27  &    4.37  &    0.62 \\
   34  &    2.66  &    3.85  &    0.57 \\
   35  &    0.52  &    2.19  &    0.60 \\
   36  &    1.69  &    3.16  &    0.65 \\
   37  &    1.97  &    2.70  &    1.22 \\
   38  &    1.10  &    2.75  &    0.64 \\
   39  &    1.07  &    2.51  &    0.83 \\
   40  &    9.23  &    5.77  &    0.59 \\
   41  &    1.08  &    2.34  &    1.03 \\
   42  &    2.79  &    2.96  &    1.31 \\
   43  &    0.85  &    2.74  &    0.51 \\
   44  &    0.50  &    2.17  &    0.60 \\
   45  &    0.64  &    2.47  &    0.52 \\
   46  &    0.28  &    1.64  &    0.79 \\
   47  &    1.41  &    2.93  &    0.68 \\
   48  &    4.18  &    4.80  &    0.46 \\
   49  &    0.85  &    2.68  &    0.54 \\
   50  &    0.40  &    1.94  &    0.67 \\
   51  &    1.20  &    2.87  &    0.62 \\
   52  &    0.24  &    1.54  &    0.82 \\
   53  &    2.13  &    4.09  &    0.38 \\
   54  &    1.40  &    3.03  &    0.61 \\
   55  &    0.30  &    1.67  &    0.80 \\
   56  &    5.18  &    4.74  &    0.59 \\
   57  &    0.31  &    1.60  &    0.92 \\
   58  &    0.79  &    2.74  &    0.47 \\
   59  &    0.37  &    1.85  &    0.72 \\
   60  &    0.43  &    2.02  &    0.63 \\
   61  &    2.60  &    3.72  &    0.62 \\
   62  &    2.30  &    3.72  &    0.55 \\
   63  &    0.36  &    1.70  &    0.89 \\
   64  &    0.41  &    1.50  &    1.47 \\
   65  &    0.72  &    1.99  &    1.12 \\
   66  &    0.98  &    2.34  &    0.94 \\
   67  &    2.84  &    4.56  &    0.37 \\
   68  &    0.38  &    1.82  &    0.76 \\
   69  &    1.98  &    3.43  &    0.60 \\
   70  &    1.14  &    2.31  &    1.12 \\
   71  &    0.42  &    1.79  &    0.89 \\
   72  &    0.72  &    2.17  &    0.86 \\
   73  &    0.68  &    2.47  &    0.55 \\
   74  &    2.96  &    4.06  &    0.54 \\
   75  &    0.25  &    1.50  &    0.90 \\
   76  &    0.52  &    2.19  &    0.60 \\
   77  &    0.40  &    1.91  &    0.71 \\
   78  &    0.34  &    1.85  &    0.65 \\
   79  &    1.49  &    3.10  &    0.61 \\
   80  &    3.21  &    4.57  &    0.41 \\
   81  &    0.43  &    1.88  &    0.78 \\
   82  &    0.44  &    1.99  &    0.68 \\
   83  &    0.76  &    2.38  &    0.68 \\
   84  &    0.49  &    2.02  &    0.73 \\
   85  &    0.66  &    2.49  &    0.52 \\
   86  &    1.12  &    2.42  &    0.96 \\
   87  &   10.29  &    5.24  &    0.87 \\
   88  &    2.25  &    3.63  &    0.57 \\
   89  &    1.36  &    2.62  &    0.93 \\
   90  &    0.49  &    2.02  &    0.72 \\
   91  &    1.09  &    2.19  &    1.26 \\
   92  &    1.61  &    2.74  &    0.95 \\
   93  &    3.55  &    3.66  &    0.88 \\
   94  &    1.06  &    2.51  &    0.82 \\
   95  &    0.70  &    1.97  &    1.12 \\
   96  &    1.11  &    2.45  &    0.92 \\
   97  &    5.86  &    5.68  &    0.39 \\
   98  &    1.34  &    2.72  &    0.81 \\
   99  &    2.22  &    3.10  &    0.91 \\
  100  &    0.61  &    2.31  &    0.60 \\
  101  &    1.87  &    2.58  &    1.34 \\
  102  &    6.71  &    5.79  &    0.42 \\
  103  &    0.27  &    1.54  &    0.89 \\
  104  &    0.53  &    2.09  &    0.71 \\
  105  &    1.64  &    2.89  &    0.83 \\
  106  &    0.83  &    2.02  &    1.24 \\
  107  &    0.46  &    1.94  &    0.77 \\
  108  &    0.78  &    2.49  &    0.62 \\
  109  &    3.63  &    3.72  &    0.86 \\
  110  &    0.37  &    1.76  &    0.82 \\
  111  &    0.22  &    1.50  &    0.79 \\
  112  &    1.59  &    2.38  &    1.43 \\
  113  &    2.39  &    3.07  &    1.01 \\
  114  &    1.14  &    2.95  &    0.55 \\
  115  &    0.89  &    2.58  &    0.64 \\
  116  &    1.20  &    2.70  &    0.75 \\
  117  &    0.58  &    1.97  &    0.93 \\
  118  &    0.62  &    2.07  &    0.86 \\
  119  &    0.88  &    2.56  &    0.64 \\
  120  &    0.42  &    1.88  &    0.77 \\
  121  &    2.15  &    4.25  &    0.34 \\
  122  &    1.34  &    2.89  &    0.68 \\
  123  &    1.55  &    2.96  &    0.73 \\
  124  &    0.34  &    1.82  &    0.69 \\
  125  &    0.26  &    1.64  &    0.72 \\
  126  &    1.50  &    3.35  &    0.48 \\
  127  &    1.49  &    2.89  &    0.75 \\
  128  &    0.27  &    1.50  &    0.96 \\
  129  &    0.36  &    1.88  &    0.66 \\
  130  &    0.75  &    2.22  &    0.84 \\
  131  &    2.91  &    4.07  &    0.53 \\
  132  &    4.67  &    4.76  &    0.53 \\
  133  &    1.94  &    3.66  &    0.48 \\
  134  &    2.19  &    4.24  &    0.35 \\
  135  &    0.44  &    1.97  &    0.71 \\
  136  &    1.79  &    3.63  &    0.46 \\
  137  &    0.30  &    1.60  &    0.88 \\
  138  &    0.26  &    1.64  &    0.73 \\
  139  &    1.68  &    3.53  &    0.47 \\
  140  &    1.07  &    2.87  &    0.55 \\
  141  &    0.75  &    2.42  &    0.64 \\
  142  &    0.39  &    1.79  &    0.82 \\
  143  &    0.35  &    1.76  &    0.78 \\
  144  &    0.31  &    1.57  &    0.98 \\
  145  &    0.56  &    2.09  &    0.74 \\
  146  &    0.71  &    2.31  &    0.69 \\
  147  &    0.37  &    1.79  &    0.78 \\
  148  &    0.45  &    1.91  &    0.79 \\
  149  &    0.39  &    1.76  &    0.86 \\
  150  &    0.81  &    2.51  &    0.62 \\
  151  &    1.35  &    2.91  &    0.67 \\
  152  &    0.46  &    1.94  &    0.77 \\
  153  &    0.83  &    2.66  &    0.54 \\
  154  &    0.60  &    2.09  &    0.80 \\
  155  &    2.22  &    3.94  &    0.44 \\
  156  &    0.27  &    1.57  &    0.85 \\
  157  &    0.76  &    2.22  &    0.84 \\
  158  &    1.76  &    3.35  &    0.57 \\
  159  &    0.84  &    2.75  &    0.49 \\
\enddata
\end{deluxetable}%

\end{document}